\documentclass[journal=jacsat,dvipsnames,manuscript=article]{achemso}

\usepackage[T1]{fontenc} 
\usepackage{amsmath,amsfonts,amssymb, mathtools}
\usepackage{geometry}
\usepackage{mathrsfs}
\usepackage{physics}
\usepackage{setspace}
\usepackage{bbold}
\usepackage{calc}
\usepackage{subcaption}
\usepackage{array}
\usepackage{textcomp}
\usepackage{gensymb}
\usepackage{subfiles} 
\usepackage{lipsum}  
\usepackage{hyperref} 
\usepackage{siunitx}
\usepackage{booktabs}
\usepackage{todonotes}

\usepackage{color,soul}
\usepackage{xcolor}
\usepackage{comment}
\usepackage{algorithm}
\usepackage{algpseudocode}

\usepackage{siunitx}
\usepackage{xr}

\usepackage{hyperref}
\usepackage{xr-hyper}
\usepackage[version=4]{mhchem}

\usepackage{graphicx}

\author{Wouter Botermans}
\email{wouter.botermans@imec.be}
\affiliation[imec]{imec, Kapeldreef 75, 3001 Heverlee, Belgium}
\author{Eric Beamish}
\affiliation[imec]{imec, Kapeldreef 75, 3001 Heverlee, Belgium}
\author{Matteo Cartiglia}
\affiliation[imec]{imec, Kapeldreef 75, 3001 Heverlee, Belgium}
\author{Liam Vandekerckhove}
\affiliation[imec]{imec, Kapeldreef 75, 3001 Heverlee, Belgium}
\author{Wouter Renckens}
\affiliation[imec]{imec, Kapeldreef 75, 3001 Heverlee, Belgium}
\author{Natan Biesmans}
\affiliation[imec]{imec, Kapeldreef 75, 3001 Heverlee, Belgium}
\author{Koen Ongena}
\affiliation[imec]{imec, Kapeldreef 75, 3001 Heverlee, Belgium}
\author{Wannes Peeters}
\affiliation[imec]{imec, Kapeldreef 75, 3001 Heverlee, Belgium}
\author{Pol Van Dorpe}
\affiliation[imec]{imec, Kapeldreef 75, 3001 Heverlee, Belgium}
\alsoaffiliation{Department of Physics and Astronomy, KU Leuven, Celestijnenlaan 200D, B-3001 Leuven, Belgium}
\author{Sanjin Marion}
\email{sanjin.marion@imec.be}
\affiliation[imec]{imec, Kapeldreef 75, 3001 Heverlee, Belgium}

\title[]{Precise Positional Readout of Molecular Barcode Structures using Solid-State Nanopores}

\begin{document}

\begin{abstract}
Fast and nonuniform translocation through solid-state nanopores limits both the detection of small molecular labels and their precise localization along molecular carriers. In this work we report the detection and localization performance of nucleotide-based molecular labels along double-stranded DNA scaffolds using solid-state nanopores in thin planar membranes. For small labels that are challenging to resolve individually, we introduce an anchoring strategy where readily detectable bulky labels serve as reference points to align multiple translocation events, enabling population-based detection and localization of smaller molecular features. The measured anchor positions constrain a probabilistic model of translocation velocity, identifying the most probable velocity profile for each event and enabling nonlinear trace "unwarping" for improved multi-event alignment. A complementary window-based evidence aggregation procedure accumulates weak but consistent label signatures across events, enabling detection of features that are individually masked by noise. These approaches enable robust recovery of single-dumbbell labels (DB1) on the order of 28 nucleotides and reduce mean localization errors to as low as 10 base pairs for DB3 labels and 40 base pairs for DB1 labels when averaging over multiple events. Stronger fractional DNA-associated current blockades, used as proxy for smaller pore geometries, are additionally associated with improved detection and lower localization error across membrane-based nanopore fabrication techniques. Overall, anchor-guided alignment provides a route to higher-density molecular information readout without compromising throughput via controlled translocation approaches.
\end{abstract}

\section{Introduction}

Nanopores are single-molecule sensors that rely on the Coulter principle \cite{coulter}, whereby ionic current blockades produced as biomolecules such as DNA and proteins translocate through nanoscale apertures provide measurements of location-specific molecular size. Biological nanopores are now seeing widespread use as DNA sequencers, as ratcheting enzymes conjugated to the nanopore slow the passage of biopolymers to speeds suitable for base calling. \cite{ratcheting_lieberman,base_calling_cherf,base_calling_gundlach,viterbi_timp,bonito} Solid-state nanopores, on the other hand, while amenable to wafer scale fabrication for potentially massively parallelized arrays \cite{imec_chip_fabrication}, are to date characterized by relatively fast translocation dynamics. As such, solid-state nanopores are better suited to resolving molecular feature sizes corresponding with 10s or more base pairs of DNA very quickly, having traded resolution for speed and molecular throughput. 

Various molecular schemes for solid-state nanopore readout have been proposed for applications in, for example, multiplexed analyte detection and quantification \cite{multiplexing_digital_encoding,multiplexing_nanoswitch,multiplexing_nanobait}, genome or protein fingerprinting \cite{fingerprinting_computational,fingerprinting_tio2,fingerprinting_simulation,fingerprinting_overview,fingerprinting_meller}, and molecular data storage\cite{dna_storage_kaikai_keyser,dna_storage_keyser_2,dna_storage_keyser_3}. Most such approaches rely on the readout of molecular "labels" disposed at specific locations along a linear carrier polymer. For example, the Keyser lab has demonstrated that DNA-based dumbbells, hairpins, and overhangs, as well as RNA structures and proteins conjugated to specific locations of a known sequence of scaffold DNA through the careful design of complementary oligonucleotide sequences, can be detected and quantified using ionic current measurements through glass nanocapillaries\cite{label_structures_keyser}.

While such work provides a robust and highly modular paradigm for encoding data in solid-state nanopore-friendly molecular vehicles, the dynamics of molecular transport through solid-state nanopores, particularly those formed in thin planar membranes, to date limit their performance for promised applications. In one aspect, the high speeds of translocation require high-bandwidth instrumentation, where temporal resolution typically comes at the cost of high measurement noise \cite{noise_smeets_dekker,flicker_noise_fragasso_dekker}, ultimately blunting the sensor's ability to resolve small label structures. Relatively bulky label structures can be engineered to skirt this signal-to-noise ratio (SNR) issue, but at the expense of scaffold real estate and the amount of data that can be encoded per molecule. 

Readout of encoded molecular information is further complicated by the Brownian motion, non-specific interactions of the DNA molecule with the nanopore surface, and the variability of the DNA molecular conformation during the threading of the polymer through the nanopore\cite{velocity_sakaue,velocity_fluctuations_lu,velocity_iso_flux_trumpet,velocity_uncertainty_plesa_dekker,velocity_tension_propagation_hsiao,velocity_tension_propagation_sarabadani_wanunu,velocity_kaikai_keyser_bell,velocity_profile_marty_vincent}. This effect, referred to herein as positional "jitter", diminishes the sensor’s ability to precisely determine where along a scaffold a particular label is physically disposed, or "localized", as identically positioned labels may appear in different relative positions even in successive ionic current signatures. This ultimately decreases the spatial resolution of solid-state nanopore readouts and hinders applications that would benefit from precise localization of molecular labels\cite{localization_meller}, such as genome or protein fingerprinting. Controlled translocation through glass nanocapillaries has been shown to increase spatial resolution and allow for remeasuring of the same molecule \cite{velocity_marion, averaging_radenovic, controlled_translocation_aleksandra}, but at the substantial cost of molecular throughput and potential for parallelization. Alternatives for controlled translocation through nanopores in thin planar membranes have included trapping DNA strands between coupled pores. \cite{flossing_dekker,flossing_drndic} However, coupling of the electric fields superimposes the signals caused by blockage of the individual pores and, as such, may complicate label readouts.

In this work, we quantify the detection and positional readout of nucleotide-based labels on double-stranded DNA (dsDNA) scaffolds translocating freely through planar solid-state nanopores. We first introduce an adaptive per-event peak-detection procedure biased towards false positive detection in which the detection threshold is derived directly from the noise fluctuations of each trace, eliminating the need for a user-tuned amplitude threshold. Candidate peaks are subsequently refined using a label-classification model that rejects false-positive detections and discriminates dumbbell-based labels of different sizes according to their measured features. 
Signals from resolvable labels on ruler constructs are then used to characterize the nonlinear relationship between measured temporal position and physical position along the scaffold, i.e. the positional jitter. Principal component analysis (PCA) of ruler-derived transport profiles defines a low-dimensional family of experimentally observed translocation distortions, while Wasserstein interpolation provides a continuous model of positional variability along the scaffold.

For labels that cannot be reliably resolved in individual events, bulky labels at known scaffold positions serve as internal reference points or "anchors" for orientation determination and trace alignment for improved trace averaging and population-based label detection. Conditioning the derived PCA model on the measured anchor positions yields a maximum \textit{a posteriori} (MAP) warping profile for each event, allowing nonlinear "unwarping" on a per-event basis before averaging, thereby reducing positional smearing. 
As a complementary, small-label-sensitive approach, we divide (aligned) current traces into equal-width windows, assign each window a p-value based on the probability that its strongest current excursion arises from label-free noise, and aggregate this evidence across events. 

Together, anchor-guided unwarping and window-based evidence aggregation enable the detection and localization of weak molecular labels that remain inaccessible through individual-event analysis or naïve time-normalized averaging, while retaining the throughput advantages of free translocation.

\begin{figure}
\centering
\includegraphics[width=1.0\linewidth]{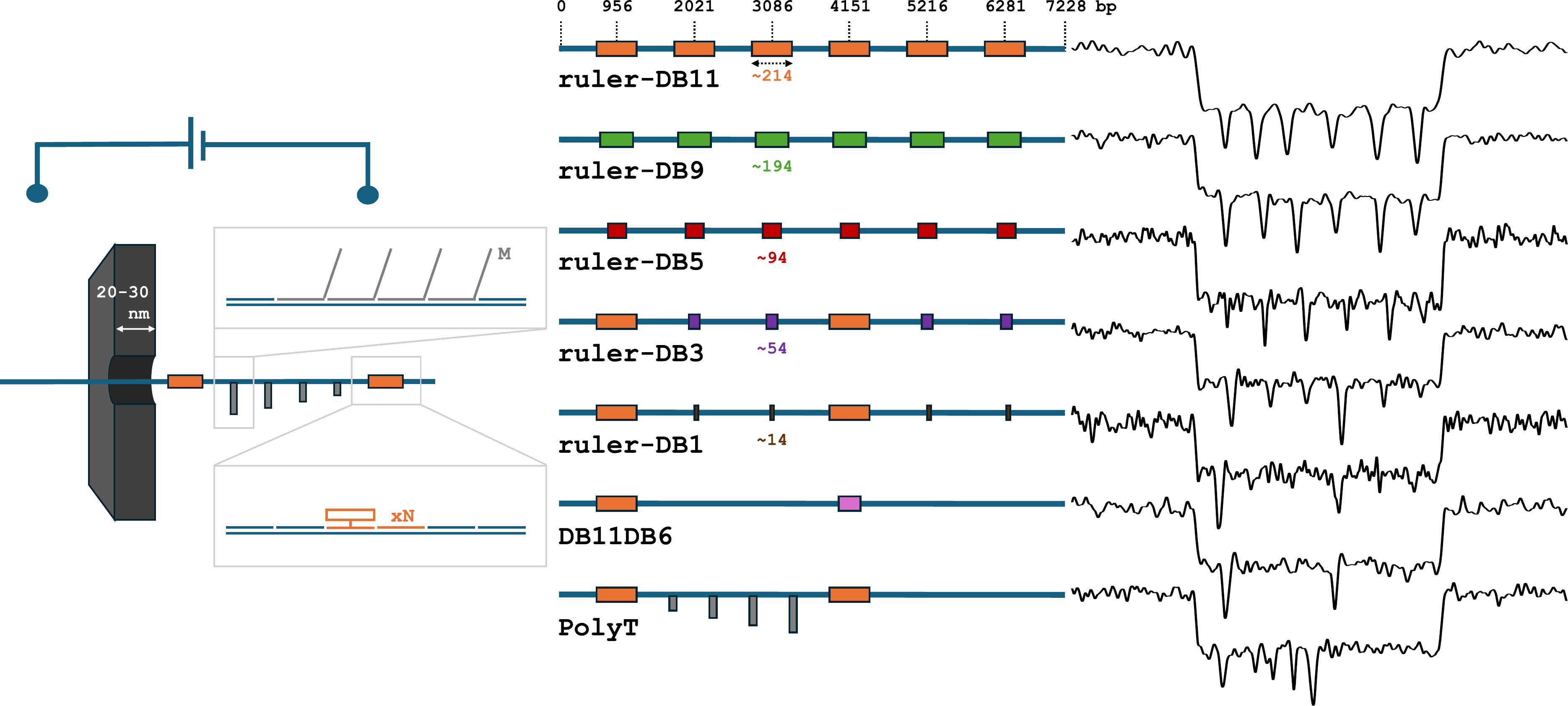}
\caption{\small \textbf{Nanopore measurements of DNA-scaffold-based labeled molecules} (left) Solid-state nanopore measurement device setup. A schematic of the dumbbell (orange) and polythymine (gray) label structures used in this paper are depicted in the insets. $\times N$ denotes the variable number of repeat dumbbell units which make up a single DB-N label structure. $M$ denotes the variable length of the four polythymine overhangs which make up a single pT-M label structure. (right) Ground truth schematic representation (not to scale) of the full molecules used in this paper along with an example trace acquired at 300 mV and low-pass filtered at 100 kHz with a 4th order Bessel filter. The center base pair locations of the ruler labels and their approximate base pair spans are annotated.}
\label{fig:overview}
\end{figure}

\section{Results}

\subsection{Single-Event Label Detection and Classification}

We begin by assessing which molecular label sizes can be detected and distinguished directly from individual nanopore translocation events. To this end, we analyzed ruler molecules carrying six dumbbell labels of defined size, spanning DB1 to DB11 (Figure \ref{fig:overview}). For the DB1 and DB3 ruler designs, two labels were replaced by larger DB11 labels to support further analysis (cf. infra). Bare label-free dsDNA served as a negative control to estimate the false-positive background.

Reliable detection of labels on single-event traces requires robust separation of label-associated changes in the current from various sources of noise. This issue is particularly important for barcode-like carriers, where even a single false-positive detected peak can change the encoded information in a molecule - especially when label positions are not known \textit{a priori}. For the detection of labels within solid-state nanopore translocation events, conventional approaches typically rely on threshold- or prominence-based peak-finding algorithms or the discrimination of state changes within the ionic current traces (e.g. CUSUM \cite{mosaic}, hidden Markov models \cite{gmmhmm}). However, these approaches typically require some sort of user-defined hyperparameter (e.g. setting a noise threshold for triggering), which may not be fully transferable across datasets and experimental conditions (e.g. due to variability in baseline noise levels) and would require operator tuning.

\begin{figure}
\centering
\includegraphics[width=1.0\linewidth]{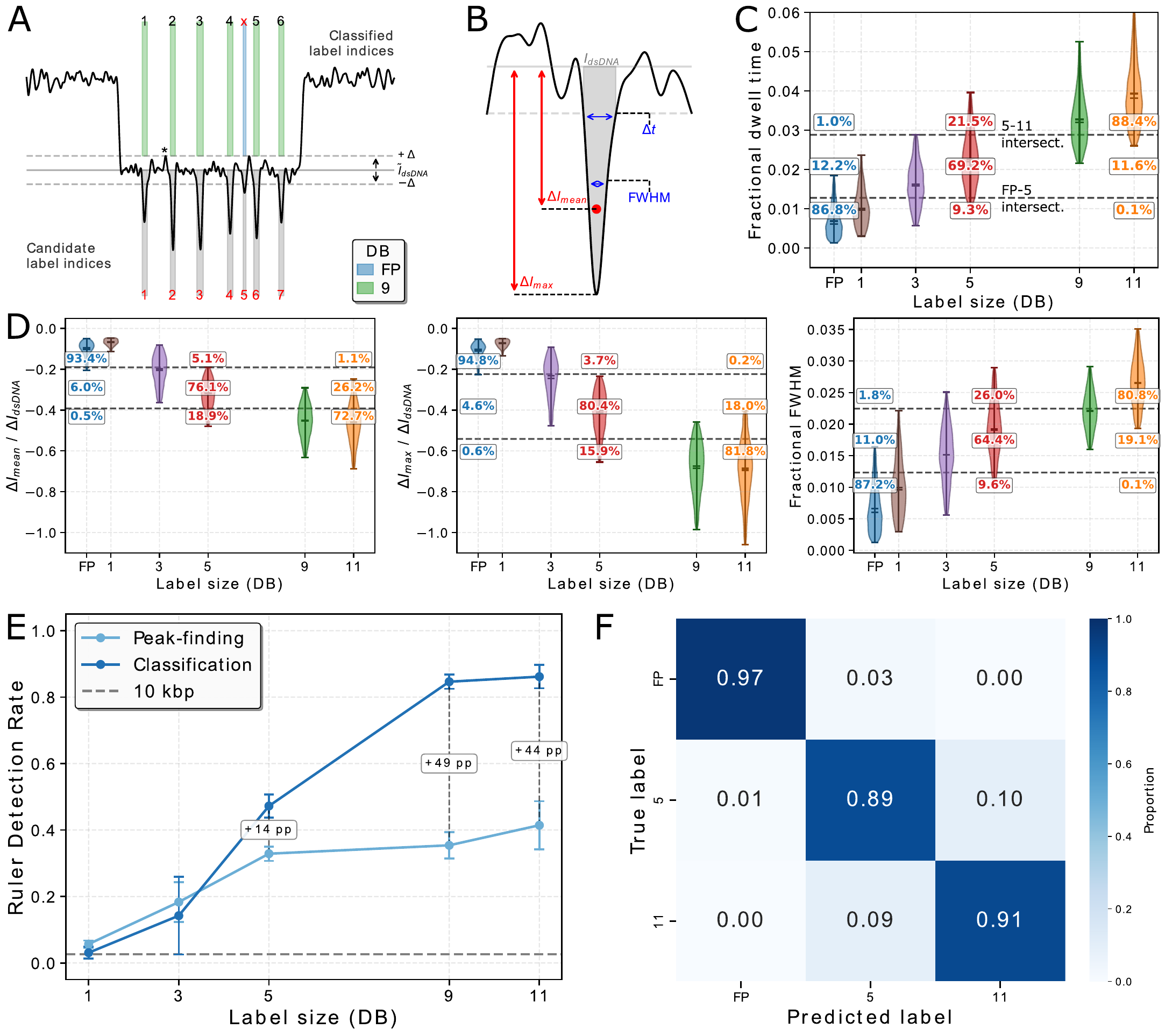}
\caption{\small \textbf{Detection and discrimination of label peaks on single molecular traces} (A) Representative translocation current trace of a DB9 ruler measured in a 10 nm nanopore at 300 mV. Gray lines indicate the dsDNA blockage level (solid) and baseline-reflected noise threshold (dashed); the asterisk marks threshold determination. Shaded regions denote detected label candidates, color-coded by predicted label size from a trained logistic regression classifier (DB9 vs false positive). (B) Zoom-in of a single label with extracted metrics (dwell time and full width at half-maximum, horizontal arrows; average and maximum depth, vertical arrows) (C) Distributions of fractional dwell time, and (D) normalized average blockage, maximum blockage, and full width at half-maximum versus label size (after removal of 5\% outliers for visualization purposes). The dashed lines mark the intersection between the distributions of false-positive labels and DB5 labels, and between DB5 labels and DB11 labels. The percentages indicate the fraction of labels per ground truth label which were classified as FP, DB5, and DB11 based on these intersections (before outlier removal). (E) Fraction of linear events with correct label count detected, comparing peak-finding alone (using the baseline-reflected noise threshold as in panel A) and classifier-refined (using a binary classification model trained on label metrics) approaches; non-overlapping mean $\pm$ SD differences are indicated by percentage point difference. The horizontal dashed line denotes the false positive baseline from bare label-free dsDNA. (F) Confusion matrix of label size predictions from k-fold cross-validation (grouped by chip) for the multiclass logistic regression model.}
\label{fig:classification}
\end{figure}

To avoid missing labels on event traces, we employed a two-step detection strategy in which the first step is a high-sensitivity pass which identifies all candidate labels, intentionally being biased towards a higher degree of false positives (cf. Methods). Assuming labels only cause negative current deflections (by design), we define the noise threshold within a current trace as the most positive deflection from the estimated baseline and reflect this threshold along the baseline to get a threshold for subsequent peak-finding (cf. Figure \ref{fig:classification}a). Any consecutive data points below this threshold are merged into a single label candidate. This per-trace, hyperparameter-free threshold preserves sensitivity while remaining robust to noise variability across events. A subsequent high-specificity classification step then rejects spurious detections based on extracted features (cf. infra).

For each detected label candidate, a set of features was extracted from the associated current drop: the fractional dwell time $\Delta t$, the mean and maximum current deviations $\Delta I_{mean}$ and $\Delta I_{max}$ relative to the open-pore-to-dsDNA current drop $\Delta I_{dsDNA}$, and the fractional full width at half-maximum (FWHM) (Figure \ref{fig:classification}b). Events from ruler molecules in which exactly six label candidates were detected were annotated with their corresponding ground truth label size value (i.e. 1-11), while any candidates detected in bare dsDNA events were annotated as false positives. For the smaller (DB1 and DB3) rulers, only the subset of events with six detected labels where the largest area-under-the-curve or AUC (as a marker of bulkiness) corresponded to the added DB11 labels, were selected (cf. Methods). Their remaining labels were annotated with the corresponding (1 or 3) target. 
We note that because annotation requires the full expected label count to be recovered by the initial peak-finder, before individual peaks can be unambiguously assigned to ground-truth label indices, this procedure selects against events in which one or more labels are weakly resolved, and can therefore bias the annotated examples of each class toward its more readily detectable instances. This selection bias is expected to be most relevant for label sizes with low single-event detection rates (cf. infra). We nonetheless adopted this full-count criterion because translocation jitter prevents unambiguous assignment of individual peaks to specific ground-truth positions on a single-event basis.

Figure \ref{fig:classification}c and \ref{fig:classification}d show the distribution of metrics from annotated labels as a function of label size for $\approx 10$ nm diameter nanopores at a bias voltage of $300$ mV.  The current drop level results from the label size and width, the velocity of translocation, the effective thickness of the nanopore, and the bandwidth of the electrical recording and applied filters. \cite{label_structures_keyser} For the case of the dumbbell-based levels here, the linear charge density of the label stays practically the same, while the width of the label increases from about 5 nm for a single dumbbell (DB1) to about 70 nm for the eleven dumbbell (DB11) label. We see that the dwell time and FWHM of the detected peaks increase monotonically with the label length, but that there is an apparent saturation of the peak depth for DB9 and DB11 labels consistent with our use of 20 nm thick nanopores. The dashed horizontal lines on the figures indicate the intersections between the metric distributions of false-positive labels and DB5 labels, and between DB5 labels and DB11 labels. The included percentages indicate the fraction of labels per ground truth label which were classified as FP, DB5, and DB11 based on these intersections. When only considering a single label metric, we see that the maximum depth provides the highest three-way classification accuracy on false-positive and DB5 labels, which are correctly classified in 94.8\% and 80.4\% of cases, respectively. For DB11 the fractional dwell time gives the highest three-way classification accuracy at 88.4\%. 

We next evaluated single-event readout performance by quantifying how often the correct number of labels was detected within a linear translocation trace. Figure \ref{fig:classification}e shows this fraction after the initial peak-finding step (light blue). As above, we report this metric as a full-count detection rate rather than per-label accuracy, since translocation jitter precludes reliable per-peak assignment to ground-truth positions on a single-event basis. Requiring recovery of the full expected label count therefore provides a conservative but robust measure of single-event performance. This definition also places an upper bound below 100\%, as contamination (inadvertent leftovers from previous experiments on the same chip) and fragment events (erroneously cleaved scaffold DNA) all reduce the probability of recovering a complete barcode. Conversely, events in which missed true labels are offset by false positives such that the total matches the expected count are still included towards full ruler detection.

To prune false positives and distinguish label sizes, we then trained binary and multiclass logistic regression models on the extracted label features for each experimental condition. The binary models were first used to remove candidate peaks predicted to be false positives, thereby generating a refined subset of events with the correct classified label count for subsequent model training (cf. Methods). Figure \ref{fig:classification}a illustrates this process for a DB9 event, in which one out of seven candidate peaks was classified as a false positive, leaving six classified labels for annotation. The application of this two-step procedure substantially improved the full-event readout (Figure \ref{fig:classification}e, dark blue). In particular, the fraction of events with six detected labels exceeded 40\% for ruler-DB5 molecules, and detection rates for DB9 and DB11 rulers surpassed 80\%, corresponding to improvements of 40-50 percentage points from peak-finding alone. These gains arise primarily from pruning false positives that would otherwise inflate the detected label count. For smaller labels, by contrast, the initial annotated subset likely remains contaminated by false positives, which limits the effectiveness of subsequent classification. Results for additional pore diameters and bias voltages are provided in the Supplementary Information.

The ability of the extracted feature set to distinguish label classes is summarized in the confusion matrix in Figure \ref{fig:classification}f, which shows the performance of a multiclass (FP vs. DB5 vs. DB11) logistic regression model evaluated with k-fold cross-validation where all measurements from a single nanopore were held out as the test set in each fold. These results indicate that the feature distributions are sufficiently stable to support transfer across nanopores of similar size, at least for the 10 nm diameter nanopores used here for calibration. The results of the full multiclass model including DB1, DB3, and DB9 labels, which inevitably shows more limited distinctive ability, is provided in the Supplementary Information. Taken together, these results show that bulky dumbbell labels can be detected and classified directly from single translocation events, whereas smaller labels approach the practical limit of single-event analysis and require population-based detection schemes (cf. infra).

\subsection{Translocation Velocity Profiling}

Accurate localization of molecular labels along a DNA scaffold requires not only that labels be detected, but also that their temporal positions within an event be translated into physical contour positions along the molecule. However, detected labels do not appear at their expected positions along the DNA scaffold when read out in time. Their apparent displacement (referred to as "jitter") emerges from the translocation process itself, where a changing balance of Brownian motion, pore-wall interactions, and molecular configuration governs translocation dynamics throughout the event. As a result, temporal coordinates cannot be mapped linearly to contour position. To correct this, assuming common underlying forces (e.g. electrophoretic driving force, viscous drag), we used labeled ruler molecules with known ground-truth positions to learn a population-level transformation from measured event coordinates to physical scaffold positions.

\begin{figure}
\centering
\includegraphics[width=1.0\linewidth]{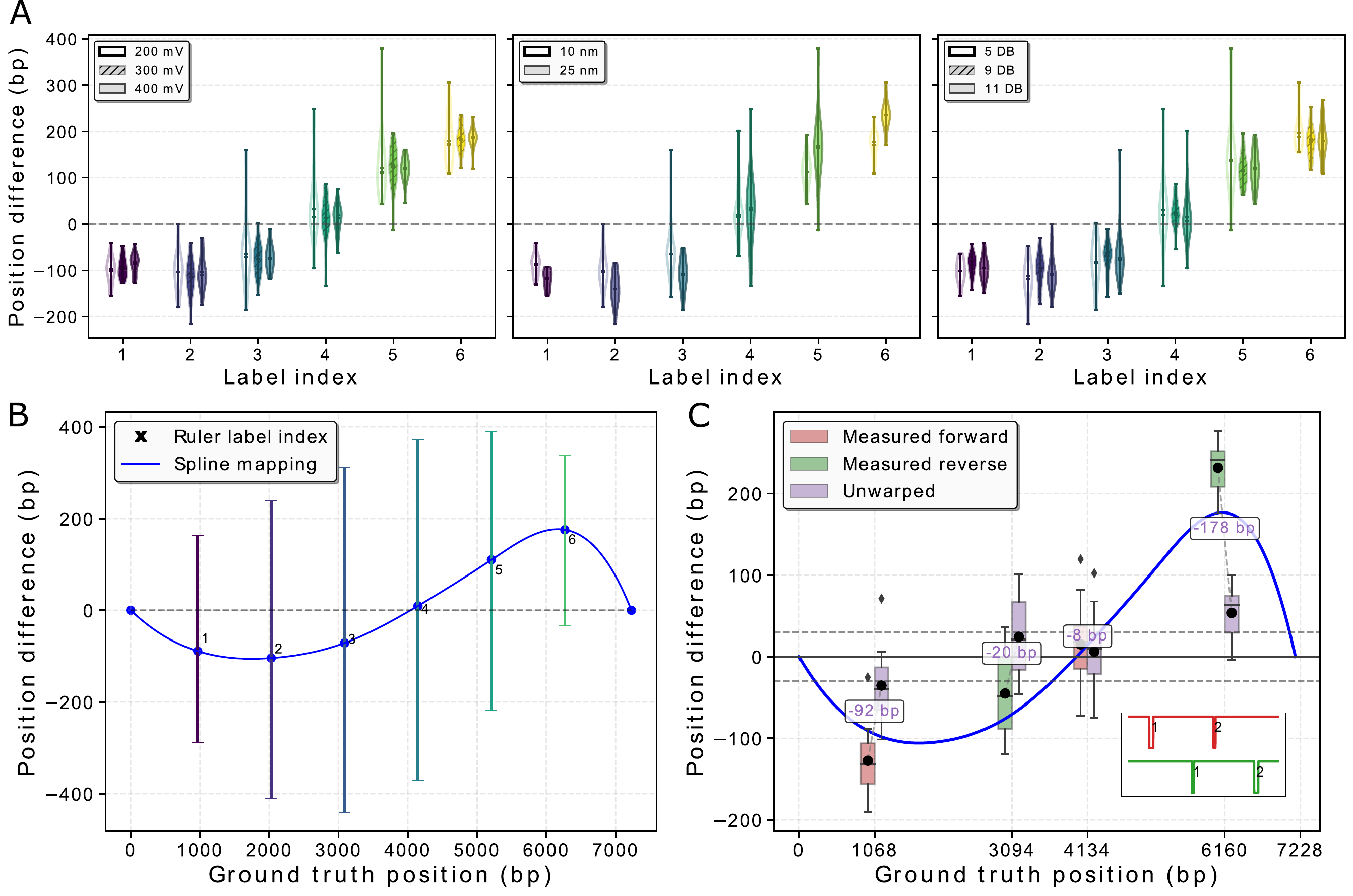}
\caption{\small \textbf{Positional shift (jitter) of label positions compared to ground truth} (A) Distribution of positional difference (measured minus ground truth base pair position) versus label index for ruler molecules under varying conditions: bias voltage (left), pore diameter (center), and label size (right). (B) Measured positional difference (mean $\pm$ SD) versus ground truth label index (1–6) for 10 nm pores; the full line represents a spline interpolation of median positions and provides a positional shift correction versus label position on the DNA strand. (C) Median positional error for forward (red) and reverse (green) translocations of a previously unseen molecule with an asymmetric DB11-DB6 dual-label construct (inset). Magenta box plots show the result of correcting for positional shift (conformational mapping) using the result from panel B. The gray dashed lines indicate the 30 bp positional difference mark.}
\label{fig:velocity}
\end{figure}

We analyzed ruler molecules carrying DB5, DB9, and DB11 labels for which all six labels had been detected in the single-event classification pipeline described in the previous section. For each event, the measured centers of the six detected labels were extracted by label index and compared with their known ground-truth positions on the scaffold, yielding six positional differences per molecule. Figure \ref{fig:velocity}a shows these positional difference distributions as a function of label index for different experimental conditions, grouped by bias voltage, pore diameter, and label size. This comparison allows us to investigate which experimental parameters materially alter the inferred velocity profile, and which can reasonably be treated as comparable for the present dataset. Across the conditions examined, the positional distributions were broadly similar when varying bias voltage and label size, whereas the nominal pore diameter produced a more visible shift in the profile shape. An analysis of pairwise mean differences supporting this trend is reported in the Supplementary Information. We therefore concluded that ruler data can be pooled across voltages and label sizes, but not across pore diameters. As such, we constructed separate mappings for the different pore-diameter classes.

The spread and shape of the distributions in Figure \ref{fig:velocity}a are consistent with known features of driven polymer transport through nanopores. Previous work has shown that non-uniform translocation, intramolecular velocity fluctuations, and end-of-translocation acceleration all complicate the direct conversion of temporal coordinates into contour positions, while more recent studies have linked the evolution of the velocity profile to polymer unfolding and tension propagation during passage through the pore. \cite{velocity_uncertainty_plesa_dekker,velocity_profile_marty_vincent,velocity_kaikai_keyser_bell} Because the effective mapping depends on the sensing geometry and experimental regime, it remains useful to determine this relationship empirically for the present system, including its pore sizes, scaffold length, label structures, and buffer conditions. The strategy adopted here is therefore not to assign absolute positions on a single-event basis, but rather to exploit the known ruler geometry to learn an average mapping that corrects the general component of this distortion.

For each pore class, we derived a continuous mapping between measured and ground-truth position by spline interpolation of the median measured position at each label index (Figure \ref{fig:velocity}b). In practice, this means that panel B combines the ruler data across the different label sizes and voltages for the 10 nm pore class and uses the median positional shift at each of the six known label sites to define a smooth correction curve. The resulting mapping shows that the positional deviation from a constant-velocity expectation approaches 200 base pairs, and that the profile is also asymmetric about the center of the scaffold. This asymmetry is modest, but it is sufficient to motivate separate treatment of forward and reverse orientations for asymmetric constructs.

The derivative of this mapping can be interpreted as an average velocity profile along the scaffold (cf. Supplementary Information). Although inferred from only six labeled positions separated by approximately 1 kbp, the resulting trend is consistent with earlier nanopore studies: translocation is faster near the ends of the molecule and more nearly constant through the middle portion. \cite{velocity_uncertainty_plesa_dekker,velocity_profile_marty_vincent,velocity_kaikai_keyser_bell} We note that this spline does not fully resolve the fine structure of the underlying dynamics, nor that it should transfer unchanged to substantially different contour lengths. Rather, it provides a coarse, experimentally grounded correction for molecules of the present scaffold length under these experimental conditions. 

An additional practical outcome of Figure \ref{fig:velocity}a is that, over the label-size range tested here, the inferred profile is comparatively insensitive to label bulk. This was not guaranteed \textit{a priori}, as attached nanostructures can in principle perturb decode performance and transport readout in barcode-style carriers, a point that has been highlighted in recent thin-membrane SiN barcode studies \cite{velocity_gevers}, although it has not been corroborated in glass nanocapillaries. \cite{label_type_influence_on_velocity_keyser} The overall consistency observed here is therefore encouraging for the ability to decouple phenomena, as it suggests that the bulky reference labels used for calibration do not dominate the transport behavior under these conditions.

Likewise, previous studies have pointed out a voltage-dependent velocity difference, with higher voltages causing more pronounced acceleration at the end of translocation \cite{velocity_profile_marty_vincent}, although these results were most pronounced outside of the voltage range which we tested here (i.e. at 100 mV and 500 mV). The same authors confirm a difference related to pore size, with larger pores having a more pronounced acceleration at the end of the translocation (cf. Supplementary Information). \cite{velocity_profile_marty_vincent}

To test the generalization of this mapping beyond the ruler constructs from which it was derived, we applied the inverse mapping to a non-ruler dsDNA molecule of the same scaffold length carrying a single DB11 and a single DB6 label in asymmetric positions along the carrier (cf. Figure \ref{fig:overview}). We selected the subset of linear translocation events for which two labels were detected after classification, and separated these into forward and reverse orientations prior to aggregation (cf. Methods). For each experiment on approximately 10 nm pores, and separately for each orientation, we then calculated the median measured positions of the first and second indexed labels and mapped these values back to scaffold coordinates using the ruler-derived calibration.

Figure \ref{fig:velocity}c compares the positional error before and after correction on the untrained construct. Without correction, the measured median label locations can deviate from their ground-truth positions by more than 200 base pairs, particularly at locations where the learned mapping differs most strongly from a constant-velocity expectation. After inverse mapping, the corrected median positions fall within 100 base pairs of the ground truth for all tested label positions, corresponding to an approximate localization precision of 30 nm or less. This represents a substantial improvement in positional readout and shows that learning an average profile from reference structures can materially sharpen localization in practical nanopore measurements. \cite{nanocapillary_localization_keyser}

The ruler-derived mapping thus identifies and corrects for a reproducible population-level component of nonuniform translocation. Applying this correction at the ensemble level, however, requires prior assignment of corresponding labels across events, which becomes difficult for densely labeled constructs or mixtures containing multiple barcode designs. Applying the mean profile directly to individual events is also insufficient as the translocation process of individual events may deviate substantially from the population average. To model this event-to-event variation, we applied principal component analysis (PCA) to spline-interpolated mappings between measured and ground-truth positions for individual ruler events. Retaining the dominant eigenfunctions defines a compact, low-dimensional parameterization of the  family of positional distortions. From this space, velocity profiles can be sampled either blindly or conditioned on specific measured label positions and their (known) ground truth positions (cf. Supplementary Information for the full derivation). 

In parallel, we constructed a continuous positional-variability model by applying one-dimensional Wasserstein geodesic interpolation to the measured-position distributions at the ruler-label locations, with boundary distributions added at the scaffold ends. For any label with known ground truth position, this allows us to infer a probability distribution of where along the trace the label will be measured. Inversely, this also allows us to infer the probability that a measured label corresponds to a predetermined ground truth label placed on the scaffold (cf. Supplementary Information).

In the next section we discuss how large labels can be used as "anchors", how we use the continuous positional-variability model to identify traces stemming from the same ground truth molecule by comparing their "anchor signatures", and how the conditionally sampled velocity profile based on these anchor positions allows for "unwarping" of the full event trace on a per-event basis.

\subsection{Anchoring-Based Label Detection and Localization}

Given the signal-to-noise ratio (SNR) challenges governing high-bandwidth solid-state nanopore measurements, robust detection of small labels on single-event traces remains difficult. In principle, averaging multiple traces originating from the same molecular ground truth should enhance SNR by suppressing random noise while reinforcing recurrent label-associated features, as has been demonstrated with controlled translocations through glass nanocapillaries. \cite{velocity_marion,averaging_radenovic,controlled_translocation_aleksandra} In the setting of free translocations through nanopores in planar membranes, however, the large inter- and intra-molecule variability in the velocity profile along the molecule ("jitter") causes substantial temporal smearing when traces are combined directly, as labels aren't time-locked to the event onset and end. Any population-based detection scheme therefore first requires a way to align traces in a physically meaningful manner.

\begin{figure}
\centering
\includegraphics[width=0.7\linewidth]{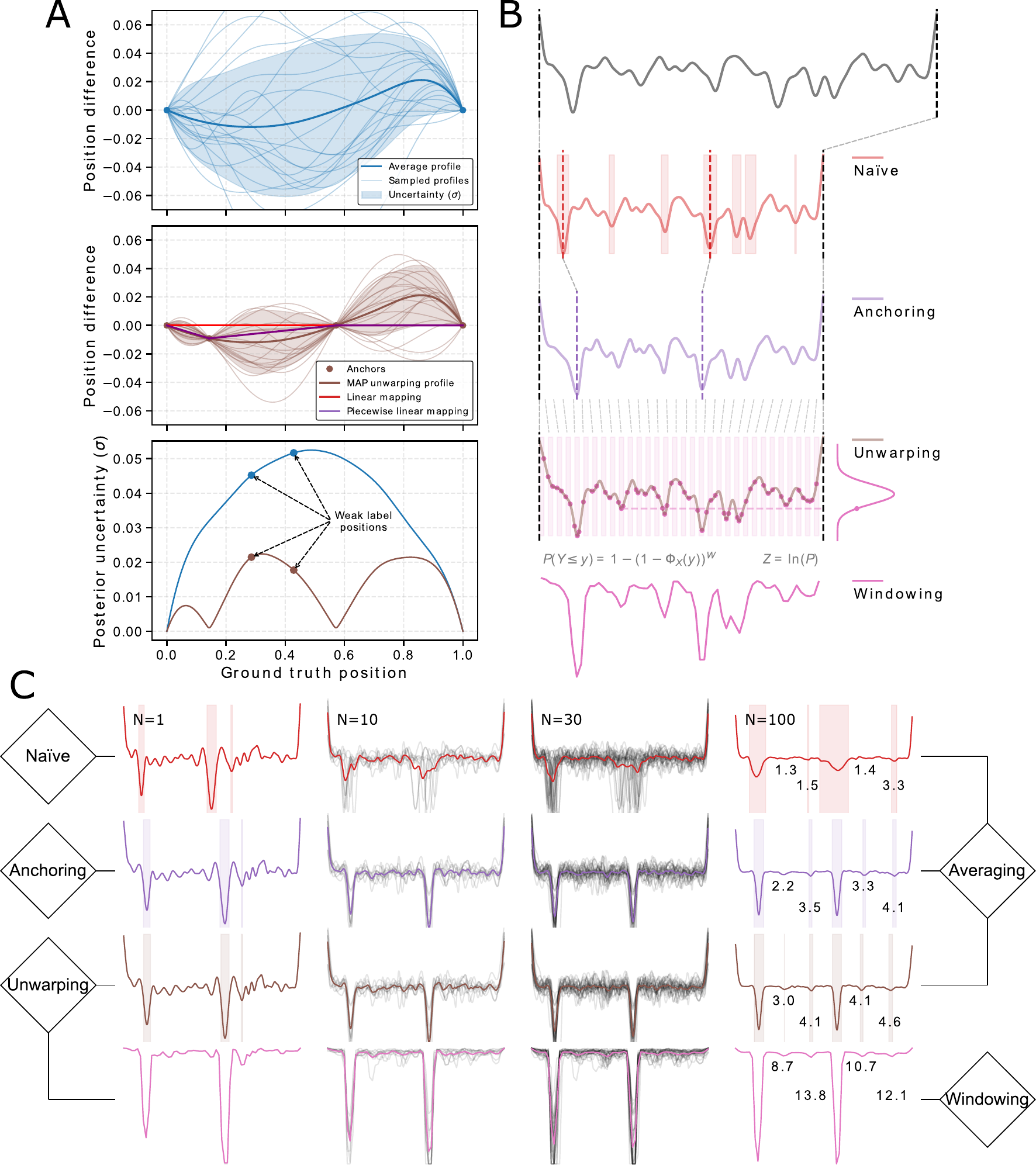}
\caption{\small \textbf{Methods for trace alignment and aggregation} (A) (top) Average measured-to-ground-truth position difference (thick line) $\pm$ SD (shaded) as a function of ground truth position, with samples from the velocity profile space (thin lines). (center) Same, conditioned on measured positions of the ruler labels with index 1 and 4 used as anchors. The red, purple, and brown lines illustrate the different alignment options, i.e. naïve, time-normalized one-to-one mapping, piecewise linear interpolation between anchors, and MAP unwarping, respectively. (bottom) SD of the position difference uncertainty from the top and center panels. The positions of the ruler labels with index 2 and 3 are annotated. (B) Illustration of the different alignment and aggregation procedures for a measured trace (black): naïve (red), anchoring (purple), and unwarping (brown) as in panel A; and windowing (pink). For the windowing approach, the equal-width windows are depicted as alternating white and pink shaded regions, with their deepest troughs (dots) compared to a Gaussian noise distribution. Multiple traces are aggregated by taking the average of the log-transposed p-values per window. The trivial case of aggregation by averaging is not shown. (C) Illustration of the aforementioned alignment and aggregation strategies applied to a sample of 1, 10, 30, and 100 (left to right) forward-oriented DB11-DB1 traces. Shaded regions indicate detected labels on the single (leftmost) and final aggregated (rightmost) traces. Label detection is not included for the windowing approach as no automatic threshold is available for this method. For each DB1 label the prominence values of the corresponding dips in the trace relative to the highest prominence values of non-label-associated dips are displayed for each method.}
\label{fig:anchoring}
\end{figure}

To address this, we introduce the concept of "anchors": bulky labels at known scaffold positions that are readily detectable on a single-event basis. These anchor labels provide internal reference points that can be used to partially "reset" the accumulated temporal jitter within a trace. The simplest implementation is an anchoring procedure in which traces stemming from the same ground truth molecule are aligned by piecewise linear interpolation between detected anchor positions and their known ground-truth locations (Figure \ref{fig:anchoring}a-b, purple). This reduces the temporal blurring that would otherwise accompany direct averaging of (time-normalized) traces and thereby improves the visibility of weaker labels in the aligned aggregate (Figure \ref{fig:anchoring}c purple versus red).
To identify traces stemming from the same ground truth molecule with known anchor positions, we can compare the set of measured anchor positions within a trace (i.e. the trace's "anchor signature") with the expected measured positions for these anchors. The expected regions can be derived from the continuous positional-variability model discussed in the previous section. Requiring all measured anchor positions to fall within a certain (central) region of the expected positional density associated with the corresponding ground-truth anchor locations, allows grouping of traces with comparable anchor signatures for subsequent alignment and aggregation (cf. Supplementary Information). This method of trace grouping also enables analysis of molecule mixtures, given that the different ground truth molecules present in the sample have sufficiently distinct anchor signatures.

Two further extensions of this idea were considered. 
First, instead of linearly interpolating between anchor segments, we can retrieve the MAP velocity profile associated with the trace by conditioning the PCA velocity profile function space derived in the previous section on the measured anchor positions within the trace. "Unwarping" the full trace in accordance with this nonlinear mapping function aims to improve localization accuracy by correcting not only the relative placement of anchor segments, but also the non-uniform transport within them (Figure \ref{fig:anchoring}a-b, brown).
To illustrate this, Figure \ref{fig:anchoring}a shows the effects of conditional velocity profile sampling. The average profile and a sample of random profiles from the PCA function space are shown, along with the standard deviation of uncertainty as a function of the scaffold position. Suppose we have prior knowledge of the ground truth positions of labels with index 1 and 4 (based on the ruler structure) on the scaffold. Using the measured positions of these labels on an individual trace allows us to select new velocity profiles conditioned on the ground truth and measured label positions. This conditioning lowers the uncertainty everywhere on the scaffold, including the positions of ruler labels 2 and 3, as shown on the figure. Furthermore, the \textit{maximum a posteriori} (MAP) conditioned profile takes into account residual nonlinearity between the outer labels, whereas simple linear interpolation between the labels would result in a skewed prediction to one side (cf. Figure \ref{fig:anchoring}a brown versus purple).

As a second extension, to emphasize detection sensitivity over positional precision, direct aggregation by point-wise averaging of the aligned traces (by anchoring or unwarping) is replaced by an evidence aggregation procedure in which each trace is divided into equal-width temporal windows and the strongest trough in each window is compared with a reference noise distribution, yielding a per-window p-value that is then combined over traces (Figure \ref{fig:anchoring}b-c, pink). 

In this formulation, unwarping favors localization accuracy, whereas windowing favors detection sensitivity. The detailed implementations of anchoring, unwarping, and windowing are described in the Methods section and in the Supplementary Information.

To illustrate the effect of these alignment and aggregation approaches towards label detection and localization, we considered ruler molecules in which two prominent DB11 labels flank a central region containing two smaller labels (DB1 or DB3) at equal spacing between the anchors (Figure \ref{fig:overview}, ruler-DB1 and ruler-DB3, respectively). This design allows the recovery of small central labels to be studied on a population basis after alignment, while keeping the anchor positions themselves readily accessible on single-event traces (Figure \ref{fig:anchoring}c). The DB11DB6 structure, which has relatively bulky labels at the anchor positions but no labels disposed therebetween, serves as a negative control for these anchor-based approaches.

Event selection for population-based analysis begins with the single-event peak detection and classification procedure described above with reference to Figure 2, extended here to a three-way classification problem (false positive FP vs. small label DB1/DB3 vs. anchor DB11). Only events in which the expected number of anchors (i.e. 2) were detected were retained for subsequent analysis. To improve robustness against outliers, and to allow for the more general case of mixture samples where several ground truth molecules carrying different anchor signatures are present concurrently, we further required the measured anchor positions to fall within 95\% of the expected positional density associated with the corresponding ground-truth anchor locations (cf. supra and Supplementary Information). 
This procedure also provides a natural basis for assigning forward and reverse translocation orientations, which were treated as separate groups throughout the population analysis.

Once grouped by anchor signature and orientation, traces were aligned by either anchoring or unwarping, zero-centered by subtracting the estimated dsDNA blockage current, and aggregated either by direct averaging or by the windowing procedure (Figure \ref{fig:anchoring}c). For label detection on aggregated traces, a noise threshold was defined analogously to the single-event case by reflecting the largest positive fluctuation about the zero line (Figure \ref{fig:classification}a). For the windowing approach, a threshold was set based on the experiments on DB11DB6 molecules under the same conditions (cf. Methods).

\begin{figure}
\centering
\includegraphics[width=1.0\linewidth]{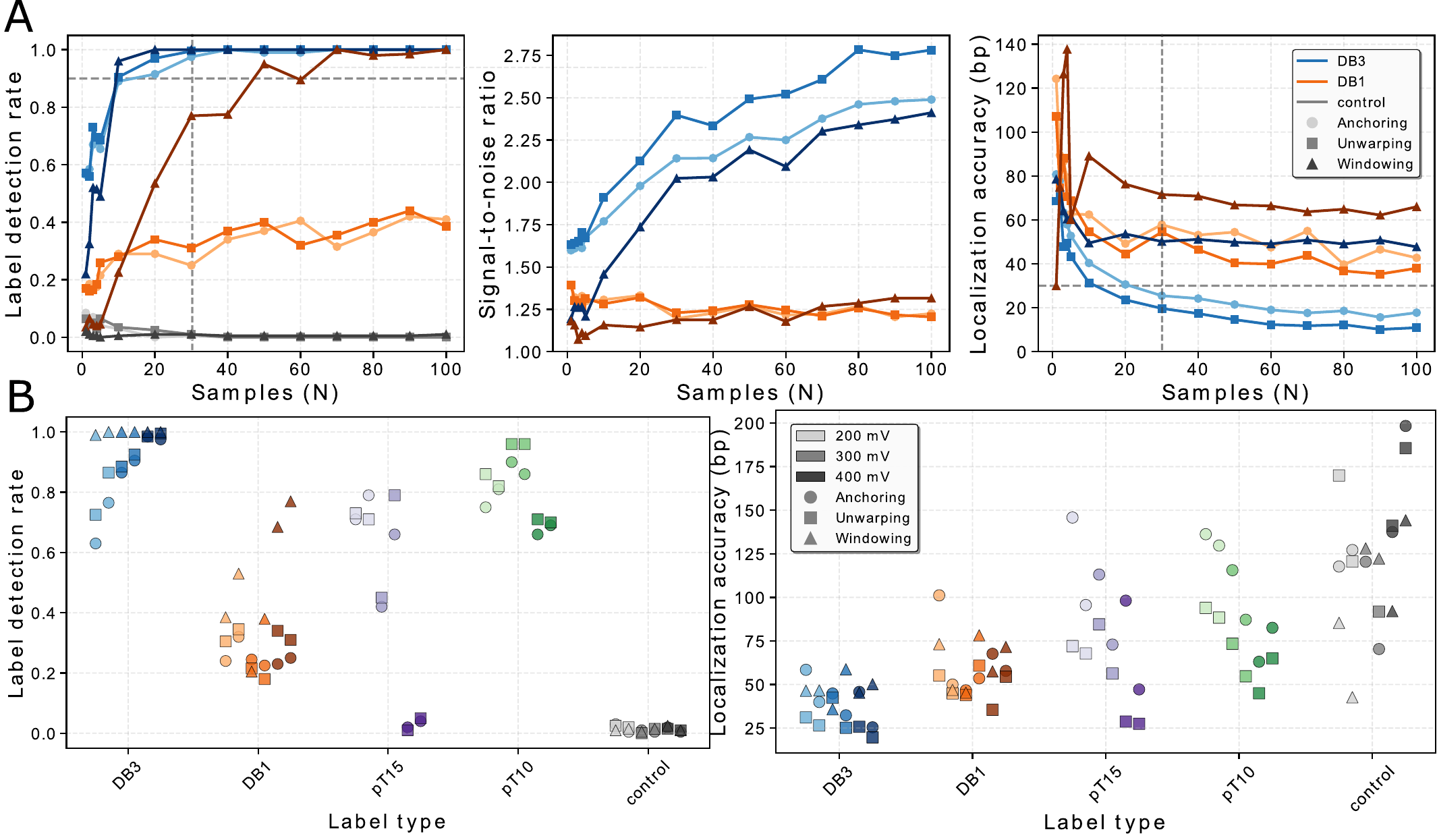}
\caption{\small \textbf{Label detection rate and localization accuracy on aggregated traces} (A) Detection rate (left), SNR (center), and average base pair localization distance from ground truth (right) as a function of the subset size for population-wise label detection of DB1 and DB3 labels between anchors. Averages over 100 iterations of a Monte Carlo simulation are reported. The DB11DB6 molecule serves as a negative control. Horizontal gray dashed lines mark the 90\% detection rate and 30 bp localization accuracy points. Vertical gray dashed lines mark the 30 events point, which correspond to the results used in panel B. The anchoring (circle, light), unwarping (square, intermediate), and windowing (triangle, dark) procedures are compared on the same data. (B) Detection rate (left) and localization accuracy (right) results from panel A at the 30-event intersection repeated for different bias voltages (light to dark colors ranging from 200 over 300 to 400 mV), molecule orientations (forward, reverse), and label types (DB3, DB1, pT15, and pT10).}
\label{fig:population}
\end{figure}

We evaluated the resulting population-based label detection by repeated Monte Carlo sampling of ruler-DB1 and ruler-DB3 traces. In each iteration, a subset of $N$ traces was sampled with replacement, aligned and aggregated (cf. supra), and the resulting population trace was analyzed for the presence of the expected small labels between the anchors. Figure \ref{fig:population}a reports the corresponding detection rate, SNR, and localization accuracy as a function of subset size for each approach (anchoring, unwarping, and windowing), with averages taken over 100 Monte Carlo iterations. Detection rate was defined as exact-match recovery of the expected central labels, SNR as the ratio of the mean detected label amplitude to the detection threshold, and localization accuracy as the mean base-pair distance between matched detected and ground-truth positions (cf. Methods). 

In Figure \ref{fig:population}a, we see that detection performance improves systematically with increasing subset size for both DB3 and, more modestly, DB1 labels. For DB3, the 90\% detection mark is reached after aggregation over roughly 10 events for all approaches, whereas DB1 remains more challenging under basic anchoring due to the small signal being further suppressed below threshold by residual temporal smearing. DB1 detection is, however, improved by the windowing approach.

Localization accuracy likewise improves with increasing trace count, falling below 100 bp near the 10-event regime and stabilizing below 80 bp with larger aggregates (the distance between the anchors on the scaffold being 3196 base pairs). Replacing basic anchoring with unwarping yields a marked improvement in localization precision, with distances dropping to as low as 10 bp for DB3 and 40 bp for DB1 while maintaining comparable detection rates. Windowing, by contrast, while providing an improvement in detection sensitivity upon aggregation, has sacrificed localization precision, which is limited by the choice of window number which defines the temporal/spatial resolution versus statistical robustness. In this sense, these approaches expose a practical trade-off between localization accuracy and detection sensitivity.

Figure \ref{fig:population}b summarizes these performance metrics at a representative intersection point of 30 aggregated events across different bias voltages (200, 300, and 400 mV), translocation orientations (forward, reverse), and label morphologies. In this example, in addition to the dumbbell-based DB3 and DB1 labels, we have also assessed the anchoring approaches for detecting single-stranded DNA overhang-based motifs, where pT10 and pT15 correspond  to clusters of, respectively, four 10-base and four 15-base polythymine overhangs per label (cf. Figure \ref{fig:overview}). Across these conditions, unwarping consistently outperforms default anchoring in localization accuracy while remaining broadly comparable in detection rate. Windowing again improves detection rate at the expense of spatial precision. The windowing procedure was not applied to the pT10 and pT15 labels, as these are too closely spaced for the present window-based analysis to operate reliably. The lower detection rates encountered for pT15 as compared to pT10 are attributable to the same underlying reason: closer label spacing in addition to larger labels precludes the trace from returning to baseline in-between labels. Given that a threshold-based instead of a prominence-based label detection approach (cf. Figure \ref{fig:anchoring}c) was chosen, the labels for which the in-between parts don't reach back below the threshold are not categorized as a true positive label, artificially lowering detection rates.
As expected, DB11DB6 control molecules show almost no detected labels for any approach, and those that are identified, are found at scattered positions. 

Taken together, these results show that bulky anchor labels can be used to transform population-based trace aggregation from a naïve averaging problem into a structured localization strategy. Anchoring alone already suppresses much of the temporal smearing caused by jitter, while unwarping further sharpens positional readout and windowing enhances sensitivity when detection is the primary objective. This flexibility is particularly valuable for molecular barcode applications where the optimal balance between label detectability and localization precision will depend on the encoding density and the intended readout task.


\subsection{Pore Diameter Dependence}

\begin{figure}
\centering
\includegraphics[width=1.0\linewidth]{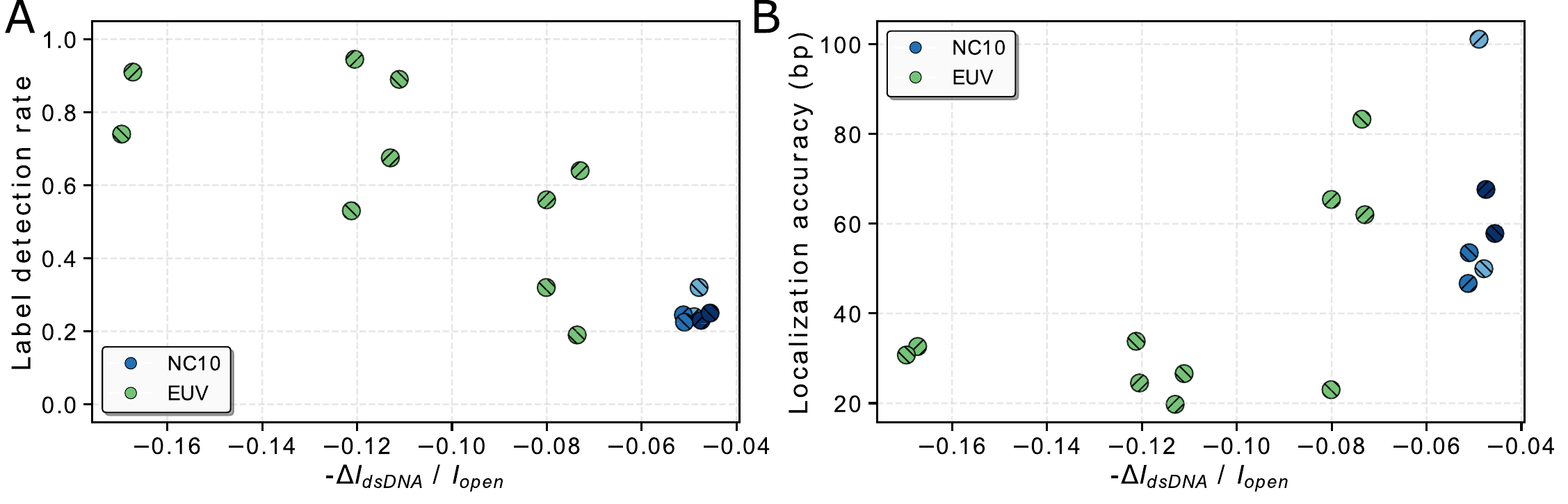}
\caption{\small \textbf{Nanopore size dependence of label localization} (A) Detection rate and (B) localization accuracy of single DB1 labels aggregated over 30 sampled events from ruler-DB1 molecules after anchoring as a function of the fractional dsDNA signal blockage as a proxy for pore size (more negative being smaller). The results shown are the averages over 100 iterations of a Monte Carlo simulation. All measurements on EUV pores were performed at a bias voltage of 300 mV and are shown per chip. For the commercial pores with a reported diameter of 10 nm (NC10), the dots are color-coded by the bias voltage (light blue at 200 mV, intermediate at 300 mV, dark blue at 400 mV), and the results are reported aggregated over the used chips. Forward/backward slashes indicate traces from events with forward/reverse orientations.}
\label{fig:in_house_chips}
\end{figure}

To assess how nanopore size influences positional readout performance, we repeated population-based analyses on a set of 300 mm wafer fabricated EUV patterned nanopores\cite{imec_chip_fabrication}, spanning estimated diameters from approximately 6 to 11 nm (cf. Methods).
All measurements shown with respect to these pores (Figure \ref{fig:in_house_chips}) were performed at 300 mV.
We repeated the DB1 anchor-based workflow described with respect to Figures \ref{fig:anchoring} and \ref{fig:population}, aggregating 30 traces to quantify detection sensitivity and localization accuracy under weak-label readout conditions as a function of pore size. 
As the EUV nanopores used for this study have a different geometry and membrane thickness (30 nm vs. 20 nm for the used commercial pores), and we expect variability in sensing region geometries between different diameters, we represent the pore diameters as the fractional current blockage caused by the passage of the dsDNA scaffold relative to the open pore current $-\frac{\Delta I_{dsDNA}}{I_{open}}$.
We only performed the anchoring-based workflow on these pores. The unwarping procedure was not used as we previously determined that velocity profiles do not carry over across different pore sizes (cf. supra). The windowing procedure was not used as this approach does not come with a "free" label threshold, and no measurements on control molecules (i.e. DB11DB6) were available for these chips.
For all analyses performed on the EUV nanopores, inference was done using models derived from calibrated data from the commercial pores. No training or fine-tuning was done on data from the EUV pores, which leads us to also study cross-manufacturing stability of our results.

Across this current blockage range we observe a clear overall trend toward improved performance with decreasing pore diameter. Smaller pores produce higher detection rates and more precise localization of DB1 labels. A change from estimated 11 nm diameter devices to 6 nm already produces a three fold increase in localization accuracy, and a large improvement in DB1 label detection. This matters not only because localization precision improves, but because reduced pore size helps move weak-label readout into a regime in which labels that are difficult to detect by conventional single-event analysis become recoverable in the first place. This also implies that when using the methods described in this paper, even smaller labels could potentially be detected on a population level.
Despite the differences between the two types of nanopore devices used, the data follow the same general performance trend, indicating cross-manufacturer stability of the anchoring approach. 

\section{Conclusion}

Solid-state nanopore readout of molecular barcodes is limited by two coupled problems: reliable identification of label-associated peaks and conversion of their temporal coordinates into physical positions along the carrier. We first address the detection problem using an adaptive, per-event peak-finding routine that derives its threshold directly from the measured noise and therefore avoids user-tuned amplitude thresholds. Subsequent false-positive refinement and label classification improve exact barcode-count recovery and distinguish larger dumbbell labels according to size. Beyond enabling automated analysis, these measurements provide a direct estimate of barcode-readout reliability, including the probability of missed and spurious labels, and can therefore inform label spacing, label design, redundancy, and error-correction requirements (cf. Supplementary Information) in nanopore-based molecular barcoding schemes.

Positional readout additionally requires correction for the nonuniform transport dynamics of freely translocating DNA. \cite{velocity_uncertainty_plesa_dekker,velocity_profile_marty_vincent,velocity_kaikai_keyser_bell} Ruler constructs with known label coordinates reveal a reproducible population-level shift between measured temporal position and physical scaffold position. Correcting this systematic distortion reduces localization errors on an independent molecular construct, demonstrating that positional interpretation requires both robust peak detection and experimentally calibrated correction for the threading dynamics of the translocating DNA carrier molecule.

Population-average correction alone, however, cannot account for the substantial event-to-event variation in translocation dynamics. We therefore introduce an anchor-conditioned maximum \textit{a posteriori} unwarping framework in which prominent labels at known scaffold positions constrain a low-dimensional family of ruler-derived transport profiles. The most probable profile consistent with the measured anchor positions is selected for each event and used to nonlinearly map the trace into scaffold coordinates before aggregation. In contrast to averaging unmodified or globally time-normalized traces, this procedure compensates for expected nonlinear transport between anchors and reduces the positional smearing of recurrent molecular features. Under the best tested conditions, conditional unwarping reduced mean localization errors to approximately 10 bp for DB3 labels and 40 bp for DB1 labels when averaging over 100 events.

We further introduce a complementary window-based evidence-aggregation scheme for cases in which detection sensitivity is more important than exact localization. Rather than averaging current amplitudes directly, the strongest current excursion in each of $N_w$ equal-width temporal windows is assigned a p-value under a label-free noise model, and statistical evidence is accumulated per window across repeated events. This approach enables recovery of recurrent label-associated features that are not distinguishable reliably from noise in individual translocations. Conditional unwarping and window-based aggregation therefore address different aspects of the same problem: the former prioritizes positional accuracy, whereas the latter increases sensitivity to weak molecular features.

The ability to recover features below reliable single-event detectability broadens the design space of scaffold-based nanopore assays. The label sizes examined here, corresponding approximately to 20–80 ssDNA bases, overlap with short nucleic-acid targets and target-induced structures such as miRNAs and related oligonucleotides (cf. Supplementary Information), whose signals have often required very small pores or non-scaffold-based detection strategies because they are difficult to resolve directly. \cite{Carson2015DirectAO,Zahid2016SequencespecificRO} More generally, weak current perturbations may also arise from proteins bound at defined positions on DNA carriers, as in multiplexed aptamer- or antibody-mediated sensing schemes. \cite{Sze2017SingleMM} Positioning multiple recognition sites between robust anchors could therefore generate spatial molecular fingerprints for protein detection, pathogen identification, metagenomic classification, or sequence-variant discrimination.

The same framework could support indirect sensing formats in which target binding induces displacement or a local structural conversion of the carrier. \cite{bell_specific_2015,Beamish2017IdentifyingSI} In such applications, individual target-associated features would not need to exceed the single-event noise threshold, provided that repeated measurements can be assigned to the same molecular design through a robust anchor signature. This creates a route toward sensitive small-target readout without relying exclusively on sub-5 nm pores \cite{zvuloni_fast_2021}, while retaining the scalability of planar solid-state nanopore platforms.

Smaller sensing geometries nonetheless continue to improve performance. Across planar membrane-based nanopore fabrication techniques and manufacturers, stronger fractional DNA blockades, used here as a proxy for reduced sensing dimensions, were associated with improved weak-label detection and lower localization error. Although pore-specific transport calibration remains necessary for nonlinear unwarping, the consistent qualitative trend across the two device platforms indicates that computational alignment and aggregation and smaller pore geometries act as complementary levers.

Taken together, these results establish a hierarchical strategy for positional molecular readout with free translocations in solid-state nanopores: adaptive peak detection and false-positive pruning improve single-event barcode interpretation; ruler-derived calibration corrects reproducible positional shifts; anchor-conditioned MAP unwarping compensates for event-specific nonlinear transport before averaging; and window-wise p-value aggregation recovers weak recurrent features when sensitivity is prioritized. This combination enables more reliable decoding of molecular barcodes and extends nanopore readout toward information-dense applications in molecular data storage \cite{dna_storage_kaikai_keyser,dna_storage_keyser_2,dna_storage_keyser_3}, DNA-carrier protein sensing \cite{bell_specific_2015,Sze2017SingleMM}, and sequence-specific nucleic-acid detection \cite{singer_nanopore-based_2010}. The present implementation assumes known anchor positions and repeated measurements of the same molecular design; extending it to blind mixture decoding and fully automated high-throughput analysis represents the next step toward deployment on parallel nanopore arrays.

\section{Methods}

\subsection{Nanopore fabrication}

All calibration measurements were performed using (nominally) 10 nm and 25 nm solid-state nanopores fabricated by Norcada. In addition, in-house EUV patterned chips with a wider diameter distribution were used to assess pore-size dependence. \cite{imec_chip_fabrication} Across the full study, measurements were performed on pores spanning an estimated diameter range of 6 to 28 nm.

Pore diameters of in-house chips were estimated assuming a cylindrical geometry \cite{kowalczyk} using the average measured open pore conductance, a solution conductivity of 17 S/m (4M LiCl/10mM Tris, pH 8), and a pore length corresponding to a nominal membrane thickness of 27 nm.

\subsection{Nanopore measurements}

Chips were suspended in a 4 M LiCl buffer in which Ag/AgCl electrodes were added to the cis and trans sides. Applied bias voltages ranged from 200 mV to 400 mV. Ionic currents were recorded at a sampling rate of 26.6 MHz using a 4x10 MHz Nanopore Reader acquisition system (Elements). Data acquisition with event detection and saving of the raw translocation snapshots was performed in accordance with the Data Sieving framework described in other recent work.\cite{datasieving}

\subsection{Molecular structures}

All label-bearing analytes consist of a 7228 bp double-stranded DNA scaffold purchased from Cambridge Nucleomics. Attached dumbbell (DB-$N$) labels consist of $N$ repeats of a short duplex hairpin structure covalently attached to the scaffold backbone. Attached polythymine (pT-$M$) labels consist of four repeats of single-stranded thymine overhangs of length $M$. A bare 10 kbp double-stranded DNA scaffold was used as an unlabeled negative control (NoLimits 10000 bp DNA Fragment, ThermoFisher Scientific).

For single-event label detection and classification, ruler molecules bore six dumbbell labels of defined size, spanning DB1 to DB11. For the DB1 and DB3 ruler designs, two DB11 labels were  incorporated at the first and fourth label positions to support subsequent annotation and orientation assignment in the population-based analyses.

\subsection{Data preprocessing}

To locate translocation events within high-bandwidth current trace snapshots, all translocation events were first digitally filtered with a 4-pole Bessel low-pass filter with a cutoff frequency at 100 kHz.  

Linear events within an experiment were selected with a custom data pipeline corroborated by visual scrutiny by a human operator. Steps in the pipeline include selection of events within a stable baseline current regime, omission of false positives, fragments, and outliers by fitting the event population AUC distribution, fitting of current blockage levels with morphology classification (linear, folded, knotted, etc.) based on the fitted levels, and subsequent selection of linear events (cf. Supplementary Information). 

\subsection{Label detection and metrics extraction}

Detection of candidate labels within a linear event was performed with a peak-finding algorithm with a per-event threshold selection procedure. Assuming label sparsity and a stable (non-drifting) blockage level during the dsDNA translocation process, the current blockage of label-free dsDNA was first estimated by taking the mode of binned current amplitudes within the event. Assuming steric labels provide excursions only toward lower absolute currents, the largest increase in absolute current from the estimated dsDNA blockage state was taken as the highest noise excursion. This value was then reflected about the estimated dsDNA blockage level and used as a threshold for label detection. All consecutive data points below this threshold were aggregated into a single candidate label, from which start and end indices were determined. By construction, this procedure is permissive toward type I errors (i.e. false positives) and is intended to retain these false-positive labels for subsequent modeling and pruning.

For each detected label candidate, the following descriptive features were extracted: the label dwell time $\Delta t$ as a fraction of the total event duration, the average and maximum current deviation $\Delta I_{mean}$ and $\Delta I_{max}$ normalized by the open-pore to dsDNA translocation current ratio $\Delta I_{dsDNA}$, and the fractional full width at half-maximum (FWHM). These metrics were used for downstream annotation and classifier training.

\subsection{Label annotation and classifier training}

Labels were annotated with their corresponding target class only for subsets of events in which the expected number of labels had been detected. For ruler molecules, this corresponded to events with six detected candidate labels. For the bare-dsDNA control samples, any number of detected peaks was retained and treated as false-positive background. For DB1 and DB3 rulers, annotation was further restricted to events in which the largest-area (AUC) candidate labels matched the expected DB11 anchor positions, corresponding to the first and fourth label index positions for forward translocations or the third and sixth positions for reverse translocations. Events not satisfying these criteria were excluded from annotation.
As discussed in the Results section, this full-count requirement, needed because jitter prevents unambiguous assignment of detected peaks to ground-truth label indices, introduces a selection bias by favoring events with more readily detectable labels of each class.

To remove false positives and classify label size, binary and multiclass logistic regression models were trained on the extracted label features for each experimental condition, defined by pore diameter and bias voltage. Binary models were first used to classify candidate labels as either false positives or true labels, thereby pruning overcalled labels generated by the permissive peak-finder. Candidate labels remaining after this pruning step were used to define a refined subset of correctly counted events for subsequent classifier training. Multiclass logistic regression models were then trained to distinguish among label sizes on the basis of the same extracted feature set.

Classifier performance was evaluated using k-fold cross-validation grouped by nanopore sensor used (i.e. nanopore ID). For single-event readout performance, detection rate was defined as the fraction of linear events in which exactly the expected number of labels was recovered after classification. As stated above, this event-level criterion was used in place of per-label accuracy because translocation jitter prevents automatic assignment of individual peaks to specific ground-truth scaffold positions on a single-event basis without manual validation.

\subsection{Velocity profile inference}

To infer a population-level mapping between measured label coordinates and physical scaffold position, we analyzed ruler molecules carrying six labels at known positions along the dsDNA scaffold. For each linear translocation event in which all six labels were detected, the centers of the detected label peaks were extracted by label index and compared with their corresponding ground-truth scaffold positions, yielding six positional differences per event.

Population-level velocity mappings were inferred separately for the pore classes retained after comparison across experimental conditions (cf. infra). In doing so, we assumed that within a given class the events were governed by comparable underlying transport physics, including electrophoretic driving and hydrodynamic drag, such that an average mapping could be meaningfully estimated from the ensemble. Median measured positions at each ruler index were spline-interpolated to obtain a continuous mapping between measured and ground-truth coordinate. The derivative of this mapping was used as an estimate of the average position-dependent translocation velocity.

\subsection{Velocity profile comparison across experimental conditions}

Velocity-profile comparisons were performed across the different experimental conditions explored in this study, including bias voltage, pore diameter, and label size. Visual comparison of the measured-to-ground-truth positional differences was complemented by an analysis of pairwise mean differences to assess whether these factors introduced systematic changes in the inferred velocity profile. These analyses are reported in the Supplementary Information.

\subsection{Orientation assignment and inverse mapping}

To assess whether the inferred measured-to-ground-truth position mapping generalized beyond the ruler constructs, we analyzed an asymmetric dsDNA molecule of the same scaffold length carrying a DB11 and DB6 label at distinct positions. Linear events in which two labels were detected after classification were retained for further analysis.

Forward and reverse translocation orientations were assigned by comparing the measured label pattern with the expected binary representation of the molecule in both orientations and selecting the orientation with minimal Wasserstein distance. For each experiment and each assigned orientation, the median measured positions of the indexed labels were calculated and inverse-mapped to scaffold coordinates using the ruler-derived calibration. The resulting corrected positions were then compared with the known ground-truth label positions.

\subsection{Anchoring and anchor signatures}

For measurements on DB11-DB1 and DB11-DB3 ruler molecules, labels detected by peak-finding were classified based on a three-way logistic regression model (false positive FP vs. label DB1/DB3 vs. anchor DB11). The set of measured positions of the labels classified as (DB11) anchors is denoted as the "anchor signature" of a trace. To group traces for population-based analysis, the anchor signature of each trace was compared to the expected range for the corresponding ground truth molecule in both orientations. Traces for which all measured anchor positions fell within the 95\% density region of their expected position distribution were retained for alignment and aggregation. The expected position distributions were derived by interpolating the measured distributions of the labels on the ruler molecules per label index with 1D Wasserstein geodesic interpolation, thereby rendering a continuous "jitter profile" (cf. Supplementary Information).

\subsection{Unwarping via conditional velocity profile selection}
 
To enable nonlinear realignment of traces between anchor segments, per-trace velocity profiles inferred from ruler molecules were represented in a lower-dimensional function space. Briefly, discretized spline interpolations of these profiles were subjected to principal component analysis (PCA), and the dominant eigenfunctions were retained as a compact basis spanning the major modes of profile variation. For each trace to be unwarped, the most likely velocity profile was then selected from this parameterized space conditioned on the measured anchor positions and their known ground-truth scaffold locations. The resulting profile was used to remap the temporal axis prior to aggregation. 

Although the selected profile for any single trace need not equal the true microscopic translocation dynamics, this procedure is expected to improve alignment on average provided that the ruler-derived training set offers a representative sampling of the experimentally accessible profile space.

Full details of the PCA construction, conditioning procedure, and remapping are provided in the Supplementary Information.

\subsection{Window-based evidence aggregation}

For the windowing analysis, each aligned trace segment between anchors was subdivided into $N_w$ equal-width temporal windows. Within each window, the deepest trough value was compared with a reference distribution representing (known or inferred) label-free noise, yielding a per-window probability for the presence of a label. Corresponding window scores were then aggregated across traces to produce a population-level detection statistic. In the analyses reported with respect to Figure \ref{fig:population}, $N_w=16$ was used. 

The DB11DB6 construct, which contains no labels in the region between the anchor labels, served as a negative control for calibration of the windowing threshold. Thresholds were chosen such that the false positive detection rate in the inter-anchor region of this control was 5\% ($\alpha = 0.05$).

Full details of the window statistic, reference distribution, and evidence aggregation procedure are provided in the Supplementary Information.

\subsection{Monte Carlo sampling procedure}

For the population-based analyses, Monte Carlo sampling with replacement was performed. Average results over 100 sampling iterations are reported.
The metrics reported for the Monte Carlo sampling procedure are calculated as follows:
\begin{itemize}
\item Detection rate is defined as the relative number of true positive detected labels between anchors if and only if no false-positive labels were detected (i.e. exact-match accuracy). False positives are defined as either excess detected labels or labels that cannot be unambiguously matched to a ground-truth location. Matching between detected and ground-truth labels is performed using a Hungarian algorithm, augmented with absorbing points halfway in-between labels with implicit infinite duplications, ensuring that spurious detections are penalized rather than matched to true labels.
\item Signal-to-noise ratio (SNR) is defined with the signal given by the mean amplitude of the detected label peaks and the noise given by the predefined detection noise threshold. As label-induced peaks are strictly negative while baseline noise is (approximately) zero-centered, our definition of SNR uses a one-sided noise measure rather than a symmetric peak-to-peak noise estimate. SNR is reported only for samples without false positive detections.
\item Localization accuracy is reported as the mean base-pair distance between the Hungarian-matched estimates and ground-truth label centers.  
\end{itemize}

\begin{acknowledgement}


\end{acknowledgement}


\paragraph{Author Contributions}
W.B., M.C., E.B., and S.M. conceived the study. 
W.B. and E.B. wrote the manuscript with input from M.C. and S.M.
W.B. developed and implemented the computational framework.
K.O. and N.B. supported the experimental and computational infrastructure.
E.B., L.V., W.R., and W.P. acquired the experimental nanopore data.
P.V.D. provided resources and institutional support. 
S.M. supervised the project and provided overall project guidance. 
All authors reviewed and approved the final manuscript.
    
\paragraph{Conflicts of Interest}
The authors declare a competing interest: a patent application covering aspects of the work reported in this manuscript has been submitted. No other competing interests are declared.

\bibliography{main}

\end{document}